\documentclass[graybox]{svmult}

\usepackage{mathptmx}        
\usepackage{helvet}          
\usepackage{courier}         

\usepackage{makeidx}         
\usepackage{graphicx}        
\usepackage{multicol}        
\usepackage[bottom]{footmisc}

\usepackage{url}             

\begin{document}

{ 

\title*{~\\[2.5em]
DNS and LES of two-phase flows with cavitation\\[-4em]
{\footnotesize{
Direct and Large-Eddy Simulation IX, Springer (in press)\\[0.5em]
Keynote Lecture ERCOFTAC Workshop DLES 9\\[-1em]April 3-5, 2013, Dresden, Germany.
}}
}
\titlerunning{DNS and LES of two-phase flows with cavitation}
\author{~\\S. Hickel\\[1em]
{\footnotesize{
Institute of Aerodynamics and Fluid Mechanics, Technische Universit\"at M\"unchen, Germany.\\ 
E-mail: {sh@tum.de}}}
}
\authorrunning{S. Hickel}
\maketitle

\section{Introduction}
\label{127_sec:1}

We report on the numerical modeling and simulations of compressible two-phase flows that involve phase transition between the liquid and gaseous state of the fluid.
Cavitation refers to the process of formation and implosion of vapor cavities within the liquid phase due to a change in pressure. Vapor cavities usually form in technical devices when the liquid is accelerated such that, by Bernoulli's principle, the static pressure falls sufficiently far below the vapor saturation pressure, or when sufficiently strong expansion waves propagate into an already preconditioned low-pressure region. When these vapor bubbles are then advected into regions where the static pressure of the surrounding liquid is above the vapor-saturation pressure, sudden re-condensation occurs, which is accompanied by a rapid acceleration of the surrounding liquid. Once all vapor is consumed, the fluids inertia leads to a strong compression of the liquid towards a focal point. This "water-hammer" effect can lead to peak pressures of several thousand bars and the emission of strong shock waves. The repeated vigorous collapse of vapor cavities is a source of noise and vibrations, and can even lead to severe structural damage of solid surfaces, so-called cavitation erosion. Cavitation erosion is one of the major reasons for premature failure of technical devices involving the processing of liquids, such as marine propellers, automotive fuel injection nozzles, liquid-rocket engine cooling channels and turbo-pump inducers. On the other hand, the effects of cavitation are exploited in industrial surface-cleaning applications and in medical applications, such as the shock wave lithotripsy. 

Numerical simulations of cavitating flows must solve complex multiphysics and multiscale problems in a consistent way. Compressibility of both the liquid phase and the gaseous phase plays a decisive role. High grid resolution and very large numbers of time steps are required: Length and time scales of bubble collapses are in the range of micrometers and nanoseconds whereas the typical scales of hydrodynamic flow features in technical devices are centimeters and milliseconds. Another major challenge is the accurate simultaneous representation of the dynamics of phase interfaces, phase change, strong shock waves and turbulence at low Mach numbers in a compressible fluid.  

We numerically study cavitation on two very different scales. 
Controlling cavitation damage requires a better understanding of the underlying fundamental bubble-collapse processes. As a first step, the high-speed dynamics of cavitation bubbles is studied in well resolved simulations with a sharp-interface numerical model \cite{lauer_2012a,lauer_2012b} on a micro scale. The underlying assumption of the employed evaporation/condensation model is that phase change occurs in thermal non-equilibrium and that the associated timescale is larger than that of the wave dynamics described by the interfacial Riemann problem. 

Direct interface resolving simulations are however intractable for real world technical applications, such as turbulent flows involving cavitation clouds with millions of vapor bubbles and a wide range of time and length scales. For this reason, we developed a coarse grained model for large-eddy simulation (LES) of turbulent two-phase flows with cavitation \cite{hickel_aldm,hickel_2011}. In LES, vapor bubbles constitute sub-grid scales that have to be modeled accordingly. On the grid scale, we solve the compressible Navier-Stokes equations for a homogeneous mixture of liquid and vapor. Infinitely fast and isentropic phase change is assumed and macroscopic effects are lumped into the constitutive relations for the homogeneous cell averaged fluid. This macroscopic model is applied to realistic technical problems \cite{hickel_2011,egerer_2013} and to more generic flows to study the interaction of cavitation and turbulence \cite{egerer_2012}.

\section{Cavitation-bubble dynamics near walls}

High-speed photography gives a first insight into the bubble dynamics during the collapse \cite{lindau_2003,tomita_1986} and shows two fundamental phenomena during the non-spherical cavitation bubble collapse process: the development of high-speed jets and the release of shock-waves upon final bubble collapse. Both, the impact of shock waves and of high-speed jets on a surface can lead to material erosion. A precise determination of peak pressures at the wall and its association with the initial bubble configuration and evolution is beyond current experimental capabilities, but can be obtained from detailed numerical simulations.

\subsection{Sharp-interface non-equilibrium model}

Our sharp-interface non-equilibrium model \cite{lauer_2012a,lauer_2012b} is intended for very detailed direct numerical simulations (DNS). 
The computational domain $\Omega$ is divided by the phase interface $\Gamma(t)$ into two different sub-domains $\Omega_{l}(t)$ and $\Omega_{v}(t)$. The volume $\Omega_{l}(t)$ accounts for the region occupied by the liquid state and $\Omega_{v}(t)$ refers to the region occupied by the gaseous state. We integrate the compressible Navier-Stokes equations separately for each fluid phase on the respective sub-domain by a standard finite volume scheme on a three-dimensional Cartesian grid. For the vapor domain, the pressure is determined from an ideal gas equation of state (EOS), and for water like liquids are modeled, for example, by Tait's equation of state. 

The standard finite volume scheme is modified in grid cells that are cut by the interface $\Gamma(t)$ in such a way that only the cell volume and face fractions that belong to the respective subdomain contribute to the finite-volume flux balance. The geometry of the moving and strongly deforming liquid-vapor interface (cell-face apertures, volume fractions and interface area) is reconstructed from a level-set field, which is advected with the local fluid velocity and describes the signed distance of any point from the immersed interface \cite{fedkiw_1999}. The zero level-set contour represents the interface between the two fluids. 

The solutions for the two sub-domains are strongly coupled through an interface interaction term $\vec{X}$, that enters the finite-volume flux balances and accounts for the mass, momentum and energy transfer between liquid and vapor. We apply a thermal non-equilibrium assumption to model the phase change at the interface. That is, during the phase-change process, the pressure is in equilibrium and the temperature has a discontinuity at the phase interface. The model of Schrage \cite{schrage_1953} is used for computing the rate of evaporation and condensation, and a two-fluid Riemann problem is solved in interface-normal direction for the interface pressure. 
From the solution of the two-material Riemann problem, the interface pressure $p_I$ and the interface normal
velocity $u_I^*$ serve to compute the pressure interaction term 
\begin{equation}
 \vec{X}_v^p = -\vec{X}_l^p = \left( \begin{array}{c} 0 \\
  p_I \Delta \Gamma ~ ( \vec{n} \cdot \hat{i} ) \\
  p_I \Delta \Gamma ~ ( \vec{n} \cdot \hat{j} ) \\
  p_I \Delta \Gamma ~ ( \vec{n} \cdot \hat{k} ) \\
  p_I \Delta \Gamma ~ u_I^* )
 \end{array} \right)\;.
\end{equation}
The mass transfer term is given by
\begin{equation}
 \vec{X}_v^t = -\vec{X}_{l}^t  = \left( \begin{array}{c} \dot{m} ~ \Delta \Gamma \\ 
  \dot{m} ~ \Delta \Gamma ~ ( \vec{v} \cdot \hat{i} ) \\
  \dot{m} ~ \Delta \Gamma ~ ( \vec{v} \cdot \hat{j} ) \\
  \dot{m} ~ \Delta \Gamma ~ ( \vec{v} \cdot \hat{k} ) \\
  \dot{m} ~ \Delta \Gamma \left( e_{v} + \frac{1}{2} \left| \vec{v} \right|^2 \right) + p_I ~ \Delta q^* ~ \Delta \Gamma
\end{array} \right)\;,
\end{equation}
where $\vec{v}$ is the velocity of the liquid at the interface in case of evaporation and the velocity of the vapor in case
of condensation, respectively. $\Delta q^* = \dot{m}/\rho_{l}$ is the phase-change induced velocity and $\dot{m}$ the
phase-change rate proposed in Ref.~\cite{schrage_1953}
\begin{equation}
 \dot{m} = \frac{\lambda}{\sqrt{2 \upi R_v}} \left( \frac{p_{sat}\left(T_{l}\right)}{\sqrt{T_{l}}}
-\frac{p_v}{\sqrt{T_{v}}} \right)\;,
\end{equation}
where $R_v$ is the specific gas constant in the vapor phase, $\lambda$ is the accommodation coefficient for evaporation or
condensation, $T_v$ and $T_{l}$ are the temperatures of vapor and liquid at the phase interface, $p_v$ is the actual
vapor pressure at the interface, and $p_{sat}$ is the equilibrium saturation pressure.

\subsection{Vapor-bubble collapse simulations}

\begin{figure}[b]
\sidecaption
\vspace{-5mm}
\includegraphics[width=0.5\textwidth]{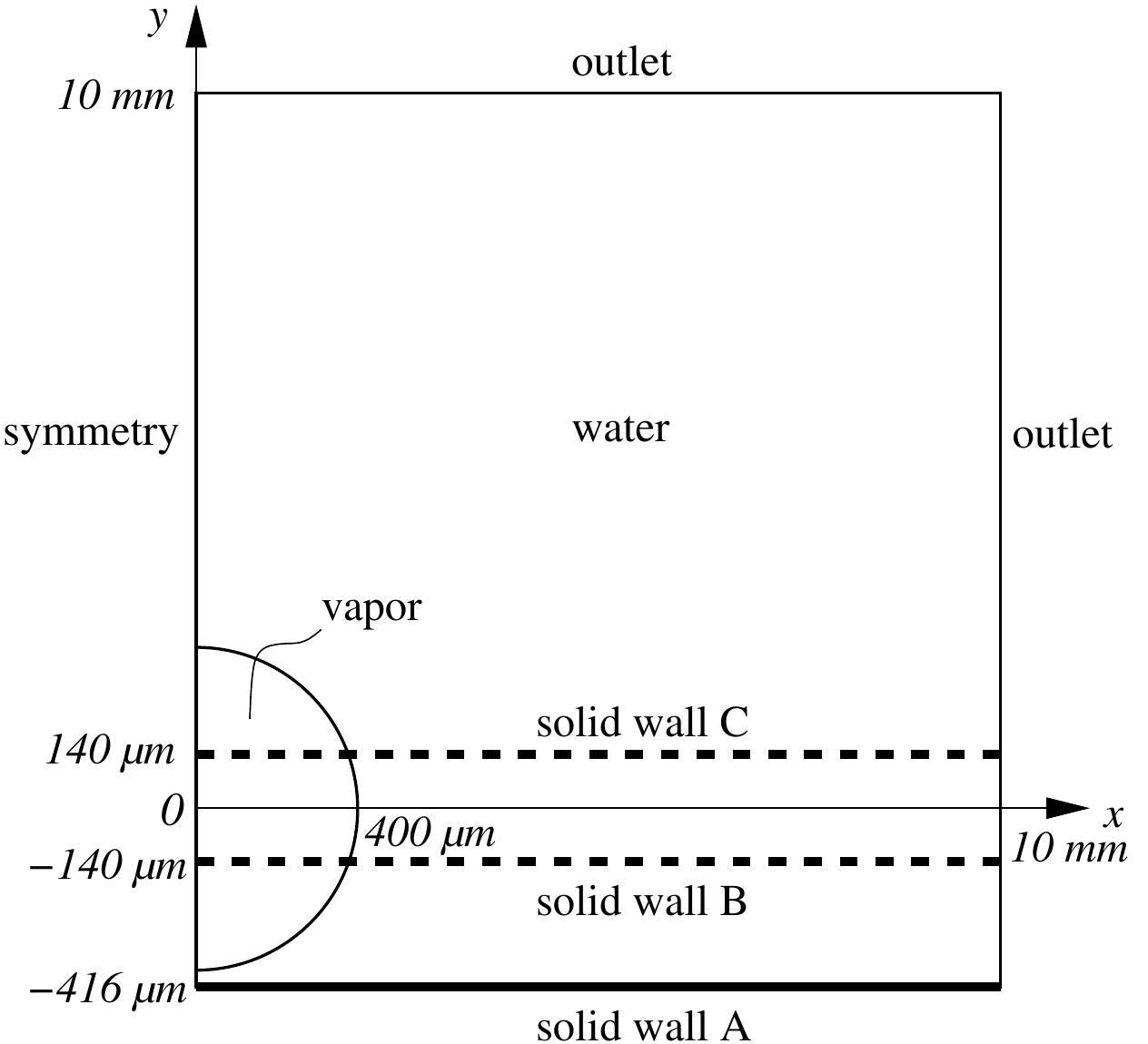}
\vspace{-3mm}
\caption{Sketch of the bubble-collapse problem. 
}
\label{127_fig:sketch}
\end{figure}

With this model, three different configurations of spherical cavitation bubbles are investigated: a detached bubble, a bubble cut by the wall in its lower hemisphere, and one cut by the wall in its upper hemisphere. 
%
As shown in Fig. \ref{127_fig:sketch}, the initial bubble radius is $400$ $\umu$m and we consider three different wall positions A, B and C. 
We take advantage of symmetries and compute only one quarter of the bubble and results are mirrored on the (X-Y)- and (Y-Z)-plane for visualization.
The grid spacing is equidistant in the bubble region with 100 computational cells over the initial bubble radius. Grid stretching is applied in the far-field. Outlet boundary conditions are imposed at $x,y,z = 10$ mm.  Both fluids have a common temperature of $293.0$\,K, which is the saturation temperature corresponding to the initial vapor pressure of $0.0234$\,bar. Initial liquid pressure is $100$\,bar and the accommodation coefficient is taken as $\lambda = 0.01$.

\begin{figure}[t]
\centering
\includegraphics[width=1.\linewidth]{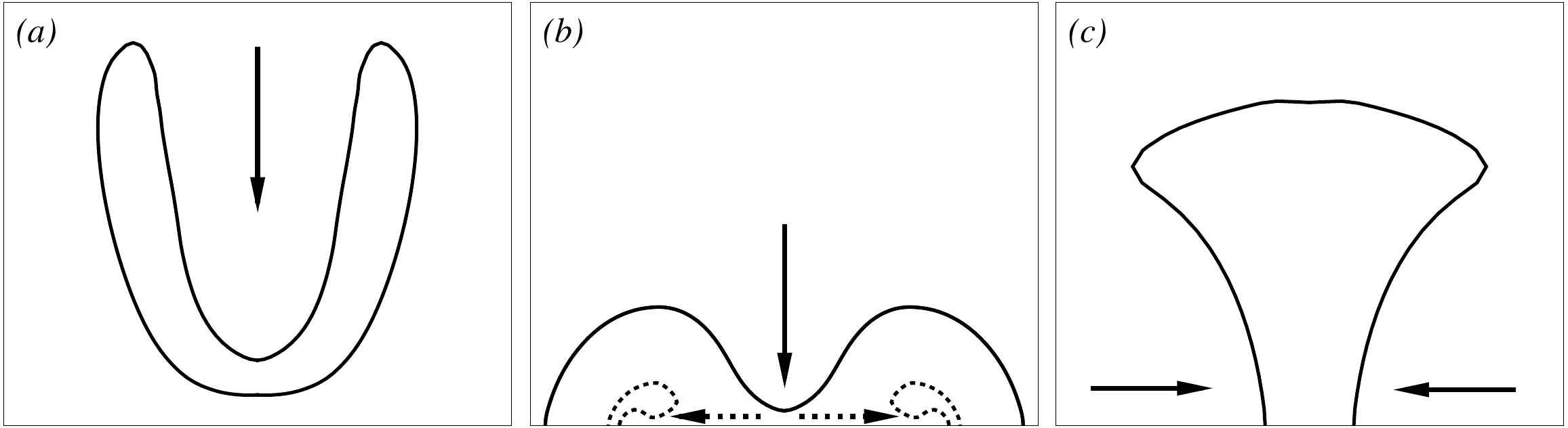}
\caption{Liquid jets during vapor bubble collapse near a wall (frame size in $\umu$m): (\textbf{a}) Wall-normal re-entrant
jet for configuration A, (\textbf{b}) primary wall-normal re-entrant jet (\textit{solid line}) and secondary
wall-parallel outward pointing jet (\textit{dashed line}) for configuration B, and (\textbf{c}) wall-parallel inward
pointing jet for configuration C.
Arrows indicate the jet direction.}
\label{127_fig:jet}
\end{figure}

\begin{figure}
\centering
\includegraphics[width=1.\linewidth]{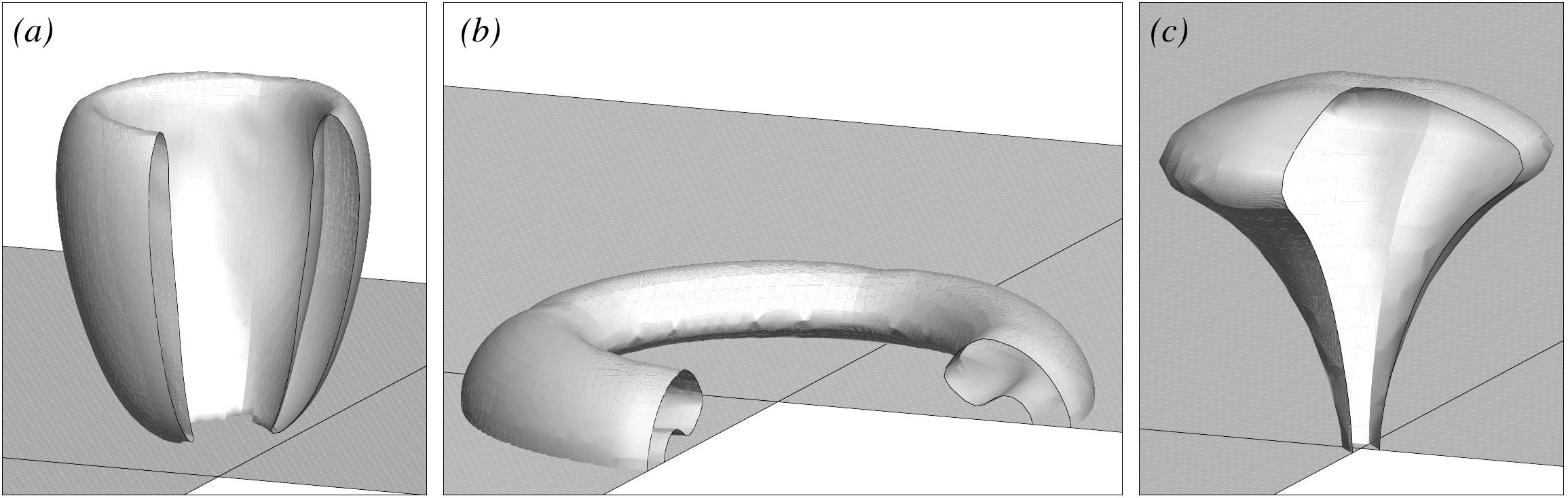}
\caption{Cuts through an iso-surface of the zero level-set (interface) showing the shape of the bubble during the late stage of the
liquid jets for configurations A, B and C.}
\label{127_fig:iso}
\end{figure}

For all configurations, the vapor bubble shrinks slowly during the initial period. The rapid stage of the bubble
collapse starts with the development of a cavity, followed by the formation of a liquid jet. Two fundamentally different scenarios
at the early stages of bubble collapse can be found. For a detached bubble or a bubble cut in the lower hemisphere, the collapse
is initiated at the top of the bubble. A fast liquid re-entrant jet develops and penetrates
through the bubble in wall-normal direction (Fig. \ref{127_fig:jet}\,a,b). For an attached bubble cut in the upper
hemisphere, the collapse is initiated between wall and interface and a liquid jet develops
radially towards the bubble center (Fig. \ref{127_fig:jet}\,c).
The appearance of a secondary jet can be observed for configuration B, where the wall normal re-entrant jet is
deflected at the wall and interacts with the remaining bubble ring (Fig. \ref{127_fig:jet}\,b, dashed lines). This
secondary jet is radially symmetric and develops from the symmetry axis outwards in wall-parallel direction. For configuration A
no secondary jet develops as the residual bubble ring is not attached to the wall.

Figure \ref{127_fig:iso} shows a three-dimensional visualization of the bubble shape during the stage where the three different
liquid jets are developed.
The first occurrence of extreme pressure magnitudes coincides with jet breakdown. For cases A and B with a wall-normal re-entrant
jet, the observed maximum wall pressures are of comparable magnitude of about $100$ times the liquid's initial pressure. Slightly larger
values for the detached bubble can be attributed to a larger jet velocity.
Looking at wall-parallel radial jets, one has to distinguish between the outward-pointing secondary jet of configuration B and
the  inward pointing primary jet of configuration C. In the latter case, the liquid is gradually compressed while being
transported towards the symmetry axis, where maximum pressure occurs. The maximum pressure after inward-pointing, wall-parallel
jet breakdown is about six times larger than that for a wall-normal jet. For the outward-running, wall-parallel secondary jet of
configuration B, very low pressure is observed inside the jet as an expansion of the liquid further decreases the pressure
of the high-velocity jet. After the jet breaks down, the liquid pressure increases, but remains significantly smaller than for the
inward-pointing jet.

During the final stage of the bubble collapse, two different scenarios occur. For cases A and C, the residual vapor bubble is
detached after jet breakdown. Thus, the maximum pressure due to final bubble collapse occurs away from the wall. The emitted
shock waves impinge on the wall with reduced magnitude, and the wall pressures do not reach the levels observed during jet breakdown.
A different scenario is observed for configuration B. After primary and secondary jet breakdown, a residual vapor ring remains at
the wall. This ring is surrounded by high pressure liquid, which initiates the final collapse radially towards the symmetry axis. The liquid
is compressed towards the center region resulting in large pressure with a maximum at the symmetry axis of about $400$ times
the initial ambient pressure.

\section{Cavitation in turbulent flows}

Coherent turbulence structures play an important role in the cavitation process in wall bounded and free shear flows. High negative pressure peaks associated with turbulent eddies can lead to cavitation inception. Vice versa, cavitation can generate and enhance vorticity, but also trigger vortex break-up. Arndt \cite{arndt_2002} gives a nice review of this complex interaction. Most technical applications involve turbulent flows for which direct numerical simulations are intractable. To address real world technical applications and fundamental questions related to the mutual interaction of turbulence and cavitation and the resulting modulation of turbulence due to phase change, we developed a coarse grained model for large-eddy simulations (LES) of turbulent two-phase flows with cavitation.

\subsection{Thermodynamic equilibrium model}

LES is based on scale separation into represented and unrepresentable scales, which is usually achieved through filtering the governing equations. The same paradigm can be applied to two phase flows. Applying a top hat filter that is consistent with a finite-volume discretization method to a cavitation cloud yields a cell averaged state vector that represents the mean density $\overline{\rho}$, mean momentum $\overline{\rho\vec{u}}$ and mean energy $\overline{\rho E}$. Accordingly, the coarse grained Navier-Stokes equations need then to be closed by constitutive relations for the mean pressure $\overline{p}$ and mean viscosity $\overline{\mu}$ of the homogenized (that is, volume-averaged) fluid.

As we consider cavitating flows with phase change, this homogenized fluid can be either pure liquid ($\alpha=1$), pure vapor ($\alpha=1$) or a mixture of liquid and gaseous state ($0<\alpha<1$) with a vapor volume fraction $\alpha$. For facilitate the reconstruction of the state within the volume of a computational cell we assume that phase changes are infinitely fast and isentropic, which is justified if the numerical timestep is large compared to the time scale of phase change. Furthermore, we assume mechanical equilibrium and we do not attempt to reconstruct the shape of the liquid-vapor interface. This means that the effect of surface tension on the vapor pressure is neglected. With these assumptions, the densities of liquid and vapor in cells that contain both phases ($ 0 < \alpha < 1 $) are $\rho_l = \rho_{l,sat}$ and $\rho_v = \rho_{v,sat}$, and we can compute the vapor volume fraction from 
\begin{equation}
   \alpha = \left\lbrace \begin{array}{lcll} 
   							0                                                                  &,& \;               &\overline\rho \geq \rho_{l,sat} \\
							\frac{\rho_{l,sat} - \overline\rho}{\rho_{l,sat} - \rho_{v,sat} }  &,& \rho_{l,sat} > &\overline\rho \geq \rho_{v,sat} \\
							1                                                                  &,& \rho_{v,sat} > &\overline\rho \\
                         \end{array} \right.
\end{equation}
Once the vapor volume fraction is known, an equation of state can be derived and the mean pressure and temperature can be computed. Note that the equilibrium pressure $p_{sat}$ and densities $\rho_{l,sat}$ and $\rho_{v,sat}$ at the saturation point generally depend on temperature, which requires an iterative procedure. An example of a barotropic equation of state (EOS) for water and water vapor is given in Ref.~\cite{hickel_2011}. Alternatively, tabulated EOS-data can be used for more complex fluids, such as Diesel fuel.  

The dynamics of wall-bounded turbulence is strongly influenced by viscous stresses. As the dynamic viscosities of liquid and vapor usually differ by several orders of magnitude, the chosen model for the viscosity of the homogenized two-phase fluid is of particular importance. In bubbly flows, the surface tension at the phase boundaries can increase the resistance against deformation substantially. For this reason we combine a straightforward linear blending of the viscosities of liquid and vapor with Einstein's \cite{einstein_1906} law for suspensions of rigid particles, which leads to
\begin{equation}
   \overline{\mu} = ( 1-\alpha ) ( 1 + 2.5 \alpha ) \mu_l + \alpha \mu_v
\end{equation}
for the volume averaged viscosity of the cell-averaged two-phase fluid \cite{beattie_1962}.

\subsection{LES of cavitation in a micro channel}

In the following we briefly report on LES of the turbulent cavitating flow in a micro-channel at two different operating points. The investigated geometry is characteristic for throttle valves found in fuel injection systems. The geometric properties of the considered micro-channel and the investigated operating points are specified in figure~\ref{127_fig:geometry}. This geometry and modifications thereof have been extensively examined experimentally at different operating conditions \cite{iben_2011,winklhofer_2001}.

The governing equations are the compressible Navier-Stokes equations for a homogeneous mixture of liquid and vapor, and we use a tabulated barotropic equation of state for the test oil Shell V1404. The system is solved by a finite-volume method on adaptive, locally refined Cartesian grids and an explicit 3rd-order accurate Runge-Kutta method for time integration. Complex geometric shapes are represented by a conservative immersed boundary method \cite{meyer_2010} that follows a similar approach as the sharp interface model for two-phase flows. The effect of non-represented turbulent flow scales (subgrid-scales, SGS) is accounted for by an implicit modeling methodology
based on the adaptive local deconvolution method (ALDM, \cite{hickel_2011,hickel_aldm,hickel_2006}), 
which yields an SGS energy transfer that is consistent with turbulence theory. 

At the inflow boundary a bi-parabolic velocity profile with bulk velocity $U_{B}$ is specified. At the outflow boundary condition we set a fixed static pressure $p_{out}$. The block structured adaptive Cartesian grid consists of roughly $31\times 10^{6}$ cells. The resolution at the wall in the micro-channel is approximately $1\times 10^{-6}$~m in order to resolve the velocity gradient at the wall.

\begin{figure}[t]
  \begin{minipage}{0.6\textwidth}
    \centering
    \includegraphics[width=\textwidth]{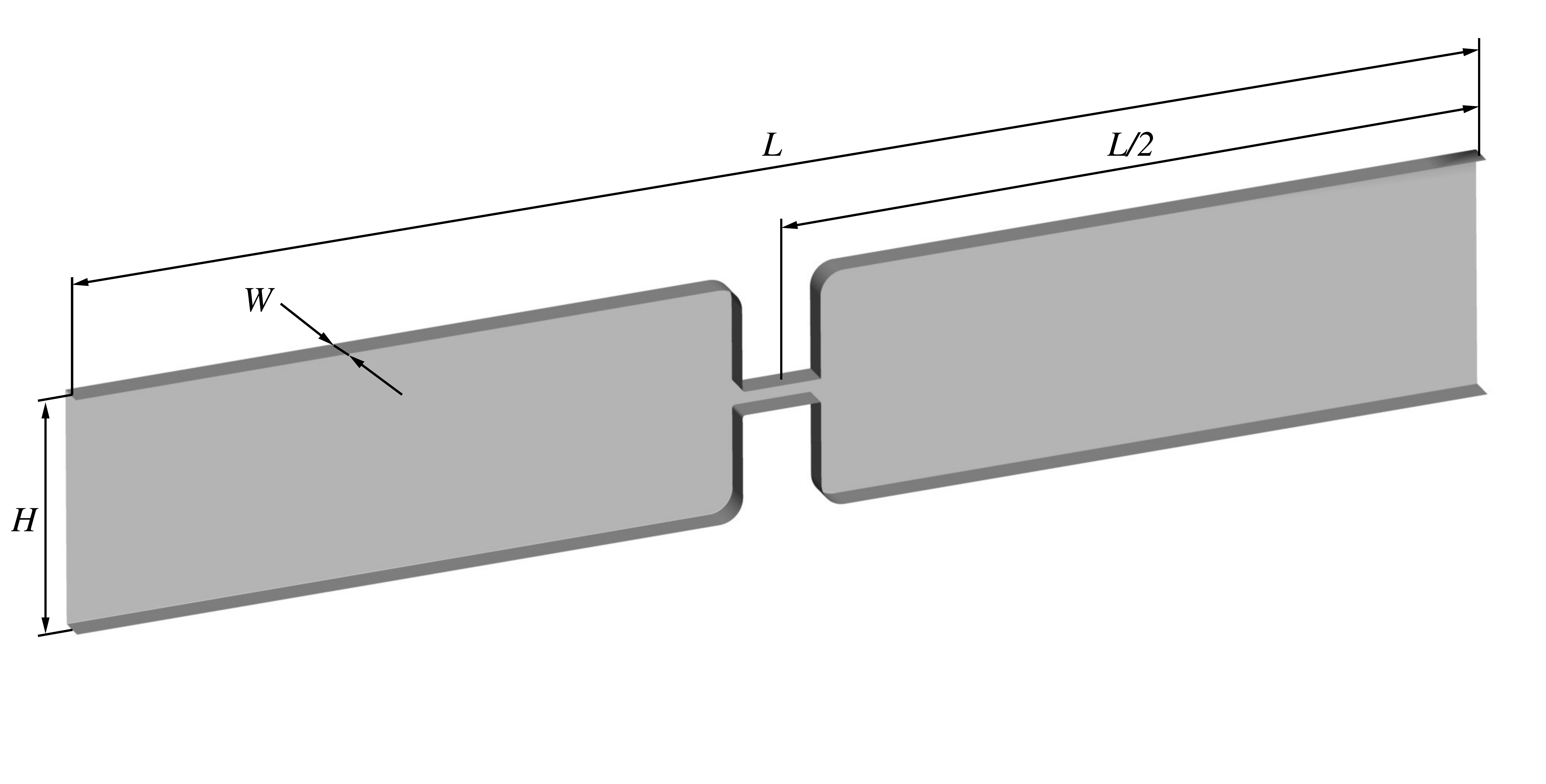}
  \end{minipage}
  \begin{minipage}{0.4\textwidth}
    \centering
    \begin{tabular}[t]{cr@{.}l}
      \hline\noalign{\smallskip}
      {Property} & \multicolumn{2}{c}{Value [$\times 10^{-3}$~m]}\\
      \noalign{\smallskip}\svhline\noalign{\smallskip}
      $L$      & 18&0\\
      $l$      &  1&0\\
      $H$      &  3&0\\
      $h$, $W$ &  0&3\\
      $R_{1}$  &  0&04\\
      $R_{2}$  &  0&4\\
      \noalign{\smallskip}\hline\noalign{\smallskip}
    \end{tabular}
  \end{minipage}\\[-6mm]
  \begin{minipage}{0.53\textwidth}
    \centering
    \includegraphics[width=0.8\textwidth]{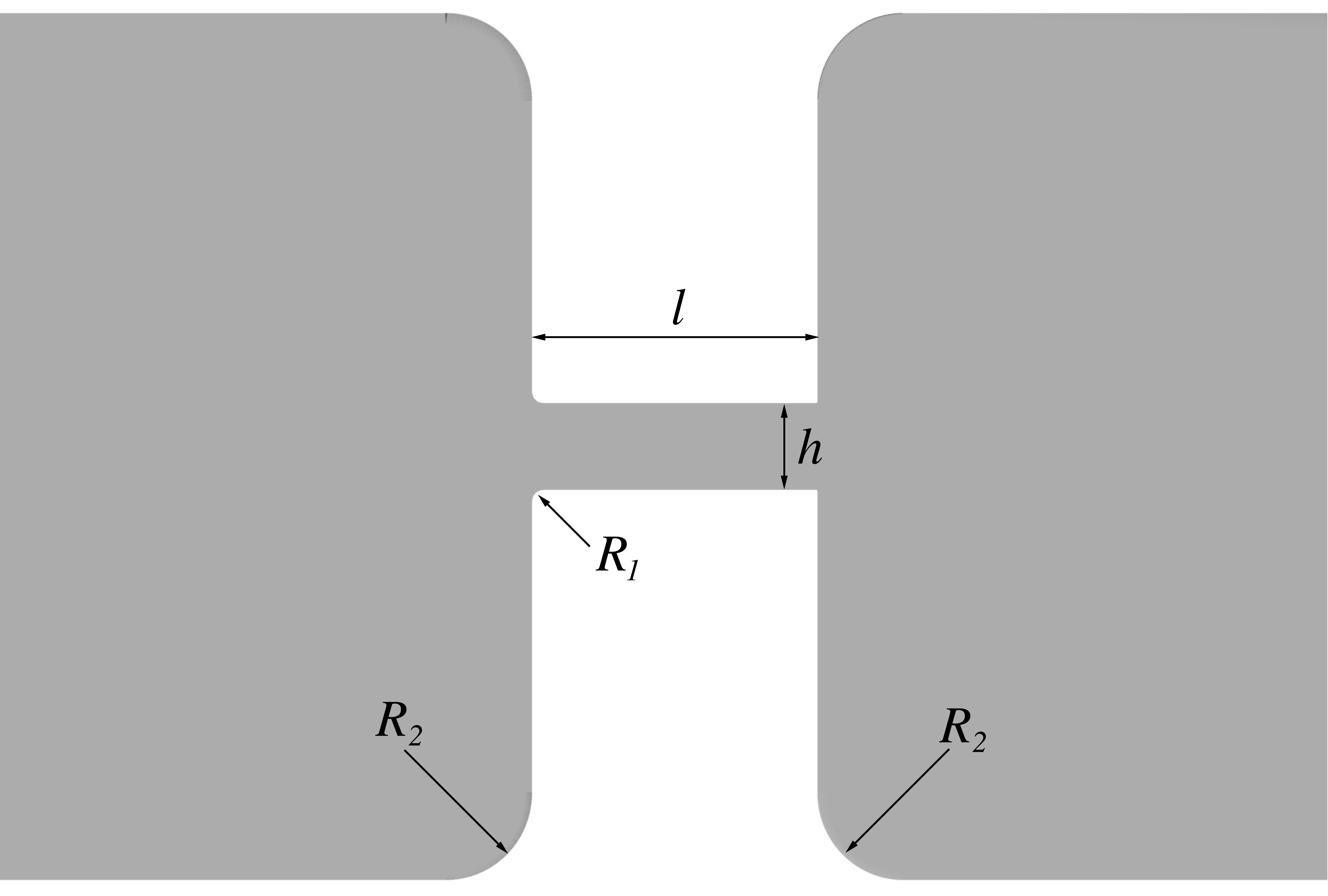}
  \end{minipage}
  \begin{minipage}[t]{0.47\textwidth}
    \centering
    \begin{tabular}[t]{crrr}
      \hline\noalign{\smallskip}
      OP & $\dot{m}$~[kg/s] & $U_{B}$~[m/s] & $p_{{out}}$~[Pa]\\
      \noalign{\smallskip}\svhline\noalign{\smallskip}
      A & 0.01450 & 19.310 & $115\times 10^{5}$\\
      B & 0.01508 & 20.078 & $55\times 10^{5}$\\
      \noalign{\smallskip}\hline\noalign{\smallskip}
    \end{tabular}
  \end{minipage}\\
  \caption{Micro-channel geometry and investigated operating points (OP).}
  \label{127_fig:geometry}
\end{figure}

\begin{figure}[h]
  \begin{minipage}[t]{0.5\textwidth}
    \centering
    \includegraphics[width=\textwidth]{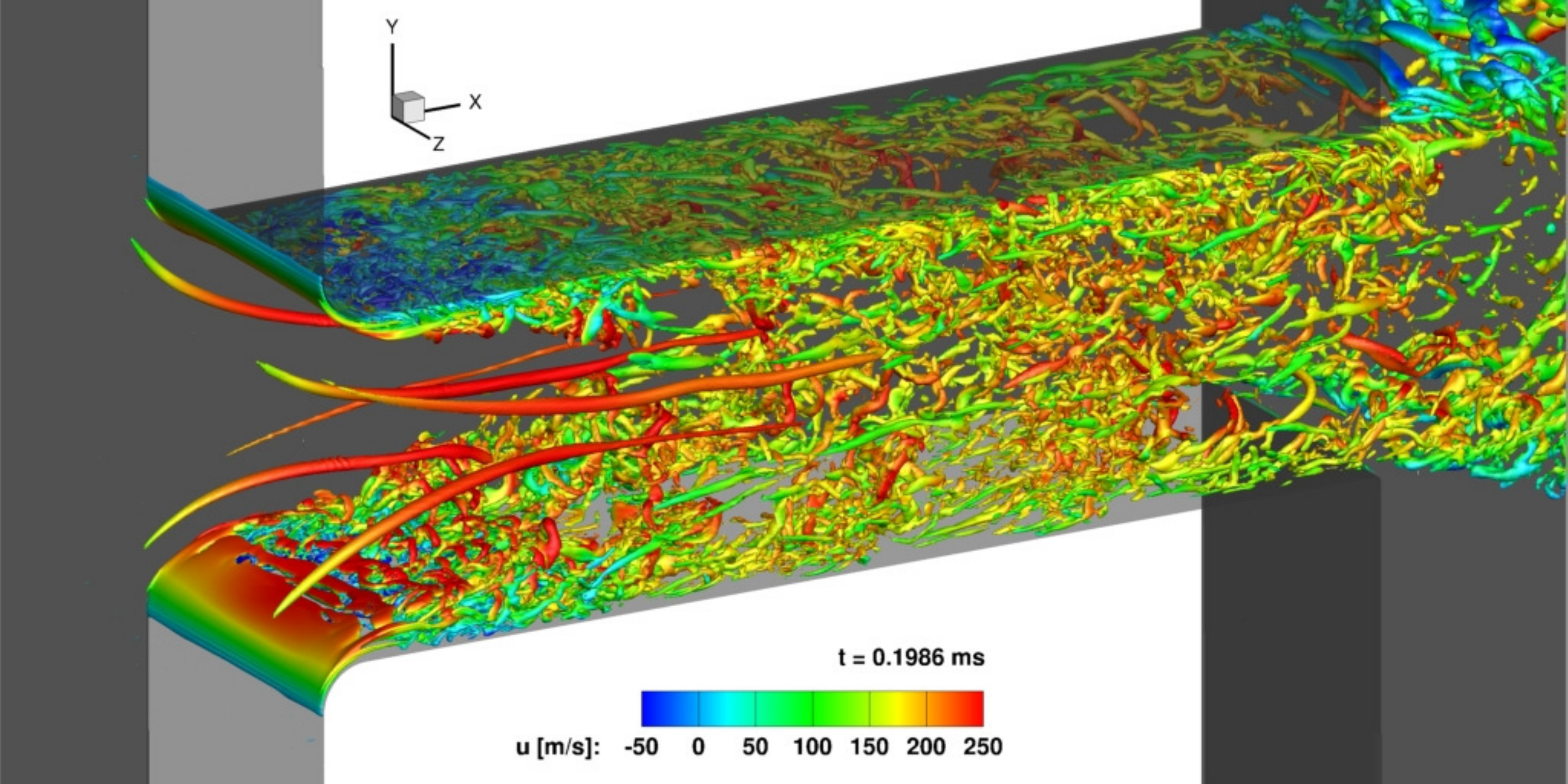}\\\vspace{0.5cm}
    \includegraphics[width=\textwidth]{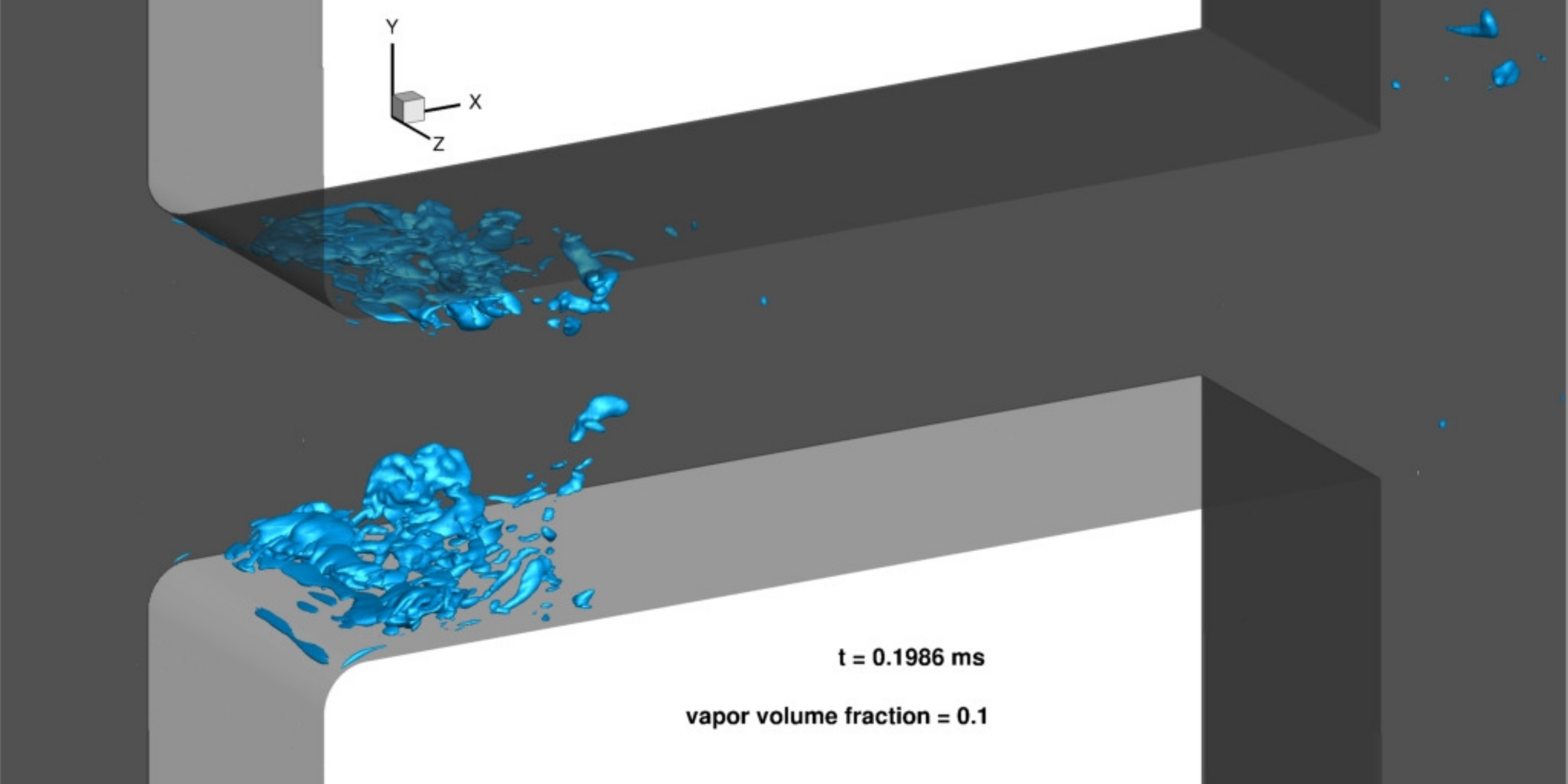}
    \\[0.5cm]{\centerline{Operating point A}}
  \end{minipage}
  \begin{minipage}[t]{0.5\textwidth}
    \centering
    \includegraphics[width=\textwidth]{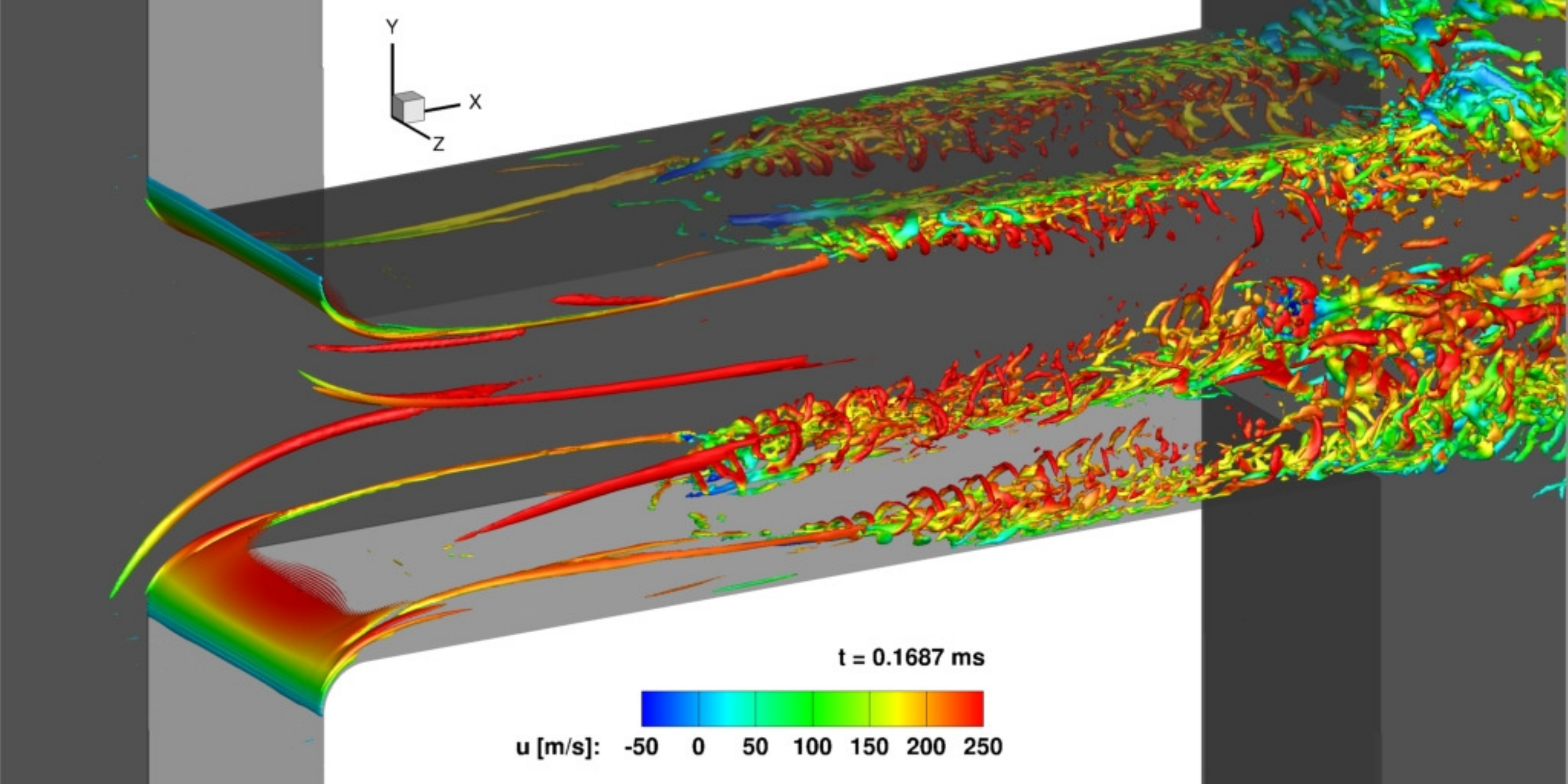}\\\vspace{0.5cm}
    \includegraphics[width=\textwidth]{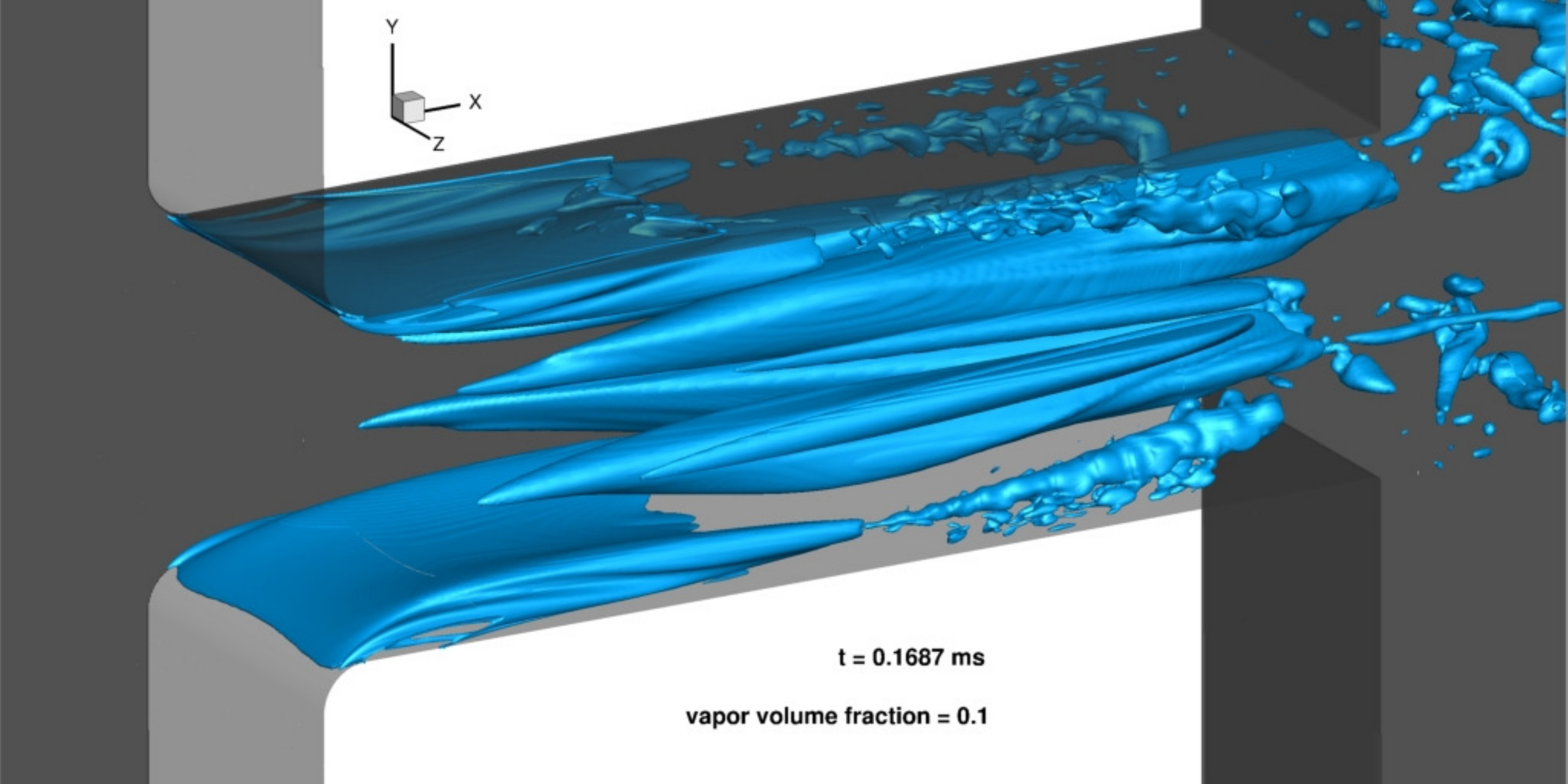}
    \\[0.5cm]{\centerline{Operating point B}}
  \end{minipage}
  \caption{Instantaneous snapshots of the flow field in a throttle valve: $\lambda_{2}$ iso-surfaces colored by the streamwise velocity (top pictures) and corresponding cavitation structures visualized by a vapor volume fraction of 10~\% (bottom pictures).}
  \label{127_fig:lambda2vapor}
\end{figure}

Figure~\ref{127_fig:lambda2vapor} shows instantaneous snapshots of $\lambda_{2}$ iso-surfaces (upper row) and iso-surfaces of 10~\% vapor volume fraction (lower row) for both operating points. These visualizations already show that phase change alters the properties of turbulence considerably.
The flow field at operating point A is highly instationary. The flow separates at the throttle inlet and forms a recirculation zone. The initially laminar duct flow undergoes sudden bypass transition within the shear layer between the bulk flow and the separation bubble. The resulting transitional and turbulent vortices are the main source of cavitation. We observe that the generated vapor cavities shed with a frequency of approximately 300~kHz.

At operating point B the back pressure has been reduced, which results in a more stationary behavior compared to operating point A. Stable sheet cavitation due to inertia effects is observed at the throttle inlet. Cavitation is also observed in the corner vortices. The $\lambda_{2}$ iso-surface indicates that the flow remains laminar throughout the first half of the throttle. Transition to turbulence is observed first at the corner vortices, which start to break up half way down the throttle valve. Furthermore, large cavitating vortex cores are present in the center of the throttle valve. These vortices originate from G\"ortler instability of the boundary layer in the chamber upstream of the throttle.

\section{Summary}


A first step towards controlling the cavitation damage requires to fully understand the underlying fundamental bubble-collapse mechanisms. For this purpose, generic configurations with spherical bubbles have been investigated by direct numerical simulations with a sharp-interface non-equilibrium model. Results for the collapse of a spherical vapor bubble close to a solid wall have been discussed for three different bubble--wall configurations. 
%
The major challenge for such numerical investigations is to accurately reproduce the dynamics of the interface between liquid and vapor during the entire collapse process, including the high-speed dynamics of the late stages, where compressibility of both phases plays a decisive role. Due to the very small timescales, liquid and vapor are in non-equilibrium at the interface, which has to be taken into account by a numerical model. Our numerical sharp-interface method copes with these problems by coupling both phases with a conservative interface-interaction term that includes a non-equilibrium phase-change model. 

Direct interface resolving simulations are intractable for real world technical applications such as turbulent flows involving cavitation clouds with millions of vapor bubbles and a wide range of time and length scales. With our thermodynamic equilibrium model for large-eddy simulations, we solve the coarse grained Navier-Stokes equations for a homogenized, cell-averaged fluid that can represent pure liquid, pure vapor and arbitrary mixtures of liquid and vapor. For deriving an appropriate equation of state, we assume that the characteristic time scale of phase change is much smaller than the numerical time step and that phase change is isentropic and in mechanical equilibrium at the saturation pressure.
This thermodynamic equilibrium model has been applied in LES of real world technical applications. Results for the cavitating flow in a throttle valve typical for fuel injection systems have been presented. We also use this model in fundamental research that contributes to the better understanding of the mutual interaction of turbulence and cavitation.


}

\end{document}